\documentclass[11pt]{article}

\usepackage[margin=1in]{geometry}
\usepackage{newtxtext,newtxmath}
\usepackage{hyperref}
\usepackage[T1]{fontenc}
\usepackage[utf8]{inputenc}
\usepackage{microtype}
\usepackage{enumitem}
\usepackage{tikz}
\usepackage{placeins}
\usepackage{graphicx}
\usetikzlibrary{arrows.meta, positioning}
\usepackage{authblk}
\title{A Governance and Evaluation Framework for Deterministic, Rule-Based Clinical Decision Support in Empiric Antibiotic Prescribing}
\author[1,2]{F.J. Gárate}
\author[1,2]{Paloma Chausa}
\author[1,2]{D. Moreno}
\author[3]{Judit López Luque}
\author[4]{Vicens Díaz-Brito}
\author[1,2,5]{Enrique J. Gómez}
\affil[1]{\textit{Grupo de Bioingeniería y Telemedicina, ETSI Telecomunicación, Centro de Tecnología Biomedica, Universidad Politécnica de Madrid, Avenida Complutense 30, 28040 Madrid, Spain}}
\affil[2]{\textit{Instituto de Investigación Hospital 12 de Octubre (imas12), Hospital Universitario 12 de Octubre, 28041 Madrid, Spain}}
\affil[3]{\textit{Innovation Unit, Parc Sanitari Sant Joan de Deu, Sant Boi de Llobregat, Spain; Institut de Recerca Sant Joan de Déu, Esplugues de Llobregat, Spain}}
\affil[4]{\textit{Infectious Diseases Department, Parc Sanitari Sant Joan de Deu, Sant Boi de Llobregat, Spain}}
\affil[5]{\textit{Centro de Investigación Biomédica en Red, Biomateriales y Nanomedicina (CIBER-BBN), Madrid, Spain}}

\date{}

\begin{document}

\maketitle

\begin{abstract}
Empiric antibiotic prescribing in high-risk clinical contexts often requires decision
making under conditions of incomplete information, where inappropriate coverage or
unjustified escalation may compromise safety and antimicrobial stewardship. While
clinical decision-support systems have been proposed to assist in this process, many
approaches lack explicit governance and evaluation mechanisms defining scope,
abstention conditions, recommendation permissibility, and expected system behavior.

This work specifies a governance and evaluation framework for deterministic clinical
decision-support systems operating under explicitly constrained scope. Deterministic
behavior is adopted to ensure that identical inputs yield identical outputs, supporting
transparency, auditability, and conservative decision support in high-risk prescribing
contexts. The framework treats governance as a first-class design component, separating
clinical decision logic from rule-based mechanisms that determine whether a
recommendation may be issued. Explicit abstention, deterministic stewardship
constraints, and exclusion rules are formalized as core constructs.

The framework defines an evaluation methodology utilizing a fixed set of synthetic,
mechanism-driven clinical cases with predefined expected behavior. This validation
process focuses on behavioral alignment with specified rules rather than clinical
effectiveness, predictive accuracy, or outcome optimization. Within this protocol,
abstention is treated as a correct and intended outcome when governance conditions are
not satisfied.

The proposed framework provides a reproducible approach for specifying, governing,
and inspecting deterministic clinical decision-support systems in empiric antibiotic
prescribing contexts where transparency, auditability, and conservative behavior are
prioritized.
\end{abstract}

\section{Introduction}

Empiric antibiotic prescribing in high-risk clinical contexts is a time-sensitive clinical
process that frequently takes place under conditions of incomplete information.
Treatment decisions are often required before microbiological confirmation is available,
and in contexts where inappropriate empiric coverage or unjustified broad-spectrum use
may compromise patient safety and antimicrobial stewardship.

Clinical decision-support systems (CDSS) have been proposed to assist clinicians in navigating this complexity and have demonstrated potential to improve care quality and safety in diverse settings \cite{Wright2010,Wright2011}. However, many existing approaches emphasize recommendation generation without explicitly specifying the conditions under which recommendations are permissible, how governance structures should be defined, or how system boundaries should be operationalized in practice \cite{Kawamanto2018}.

This work addresses these challenges by specifying a governance and evaluation
framework for deterministic, policy-based clinical decision support. Rather than seeking
to optimize predictive accuracy or clinical outcomes through learning mechanisms, the
framework prioritizes explicit specification of system behavior and its governing
constraints. The contribution is methodological: it defines a reproducible approach for
specifying, governing, and inspecting the behavior of deterministic clinical decision-support systems in narrowly scoped, high-risk prescribing scenarios.

\subsection{Contributions}

This paper makes the following contributions:

\begin{itemize}
    \item It specifies a governance layer for deterministic clinical decision-support systems, formalizing when recommendations are \emph{permissible} and when the system must \emph{abstain}.
    \item It defines core governance constructs---including explicit abstention, exclusion rules, and deterministic stewardship constraints---as first-class components separate from clinical decision logic.
    \item It proposes an evaluation protocol based on a fixed suite of synthetic, mechanism-driven cases with predefined expected behavior, enabling reproducible behavioral inspection and auditing.
\end{itemize}

\section{Scope, Design Philosophy, and Motivations}

\subsection{Constrained Scope as a Safety Feature}

This framework operates under explicitly constrained clinical scopes. Scope limitation
is treated as a deliberate design choice that enables precise specification of governance
rules and clinical logic.

System outputs are intentionally limited to two possible actions: (i) the provision of an
empiric antibiotic class recommendation, or (ii) an explicit abstention from
recommendation. Abstention is considered a correct and intended outcome when
predefined safety, governance, or information constraints are not satisfied.

\subsection{Determinism and Conservative Decision-Making}

Deterministic behavior is a foundational design principle. Given identical inputs, the
system produces identical outputs without reliance on probabilistic inference, statistical
learning, or adaptive mechanisms. This design choice prioritizes transparency,
traceability, and reproducibility over extrapolative coverage.

Conservative decision-making is enforced through explicit constraints rather than
implicit heuristics. The framework favors non-recommendation in the presence of
incomplete inputs, conflicting signals, or unresolved clinical ambiguity. No silent
generalization beyond the defined scope is permitted.

\subsection{Separation of Clinical Logic and Governance}

Clinical logic and governance mechanisms are treated as distinct design layers. Clinical
decision rules define what may be recommended under strictly defined conditions,
while governance mechanisms determine whether a recommendation is permissible in a
given context. Previous work has emphasized the importance of governance structures in ensuring safe and sustainable CDSS deployment, but has largely focused on organizational processes rather than formalized system-level constraint architectures \cite{Wright2011,Kawamanto2018}. This architectural separation supports:

\begin{itemize}
    \item \textbf{Auditability}: Each layer can be inspected independently.
    \item \textbf{Controlled policy evolution}: Clinical rules can be updated without altering
    governance structure.
    \item \textbf{Explicit constraint enforcement}: Safety boundaries are enforced structurally
    rather than heuristically.
\end{itemize}

This separation is particularly important in high-risk prescribing scenarios where the
rationale for both recommendation and abstention must be traceable to explicit rule
activation.

\subsection{Trade-offs and Design Rationale}

A number of frameworks, models, and implementation strategies have been proposed to guide the adoption and evaluation of CDSS in clinical settings \cite{Fernando2023}. However, explicit formalization of governance mechanisms as first-class architectural components remains limited. The deterministic, rule-based approach adopted in this framework reflects specific
trade-offs:

\begin{itemize}
    \item \textbf{Transparency over coverage}: The system abstains when governance conditions
    are not met, rather than attempting to extrapolate recommendations through
    statistical inference.
    \item \textbf{Explicit specification over learned behavior}: All decision logic is specified
    \textit{a priori} rather than learned from data, enabling full traceability at the cost of
    requiring domain expertise for rule definition.
    \item \textbf{Conservative behavior over recommendation maximization}: The framework
    prioritizes avoiding inappropriate recommendations over maximizing coverage.
\end{itemize}

These trade-offs are appropriate for contexts where decision support must operate
within well-defined clinical policies, where recommendation justification is required,
and where uncontrolled system behavior poses safety risks.

\section{System Overview}

The proposed framework is organized as a deterministic clinical decision-support
system in which recommendation eligibility is evaluated through an explicitly governed
sequence of logical steps. As illustrated in Figure 1, the framework separates clinical
decision logic from governance mechanisms, ensuring that recommendation eligibility
is explicitly evaluated prior to output generation.

At a high level, the system processes a clinical case by separating three conceptual
layers: clinical decision logic, governance mechanisms, and evaluation logic. Clinical
decision logic encodes the conditions under which an empiric antibiotic class may be
considered appropriate within the defined scope. Governance mechanisms act as
gatekeeping constructs that determine whether a recommendation is permissible given
the available information, safety constraints, and stewardship requirements. Evaluation
logic is external to the decision process and is used solely to inspect system behavior
against predefined expectations.

Recommendation generation follows a strictly sequential flow. Input data are first
assessed for completeness and consistency. If mandatory information is missing or
conflicting, governance rules may trigger immediate abstention. When input conditions
are permissive, clinical decision rules are evaluated to identify candidate empiric
antibiotic classes. These candidates are then subjected to stewardship and exclusion
constraints, which may further restrict or veto recommendation issuance.

The system produces one of two possible outputs: an empiric antibiotic class
recommendation or an explicit abstention. No ranking, confidence scoring, or
probabilistic output is generated. Each output is fully determined by the input data and
the active rule set, ensuring reproducibility and traceability.

Importantly, the framework does not assume that a recommendation will be generated
for every case. Non-recommendation is treated as an expected outcome when
governance conditions are not satisfied. This design explicitly avoids fallback behavior,
silent default recommendations, or implicit extrapolation beyond validated policies.

The system architecture is intentionally modular at the conceptual level. Clinical rules,
governance constraints, and evaluation procedures are specified as independent
components, allowing controlled modification of individual elements without altering
the overall decision semantics. This separation supports auditability and facilitates
systematic inspection of system behavior under different governed conditions.

\begin{figure}[t]
\centering
\begin{tikzpicture}[
    node distance=1.6cm,
    box/.style={
        draw,
        rectangle,
        rounded corners,
        align=center,
        minimum width=6.2cm,
        minimum height=1.2cm
    },
    dashedbox/.style={
        draw,
        rectangle,
        dashed,
        rounded corners,
        align=center,
        minimum width=7.2cm,
        minimum height=2.8cm
    },
    arrow/.style={->, thick}
]

\node[box, fill=blue!10] (clinical) {
\textbf{Layer 1: Clinical Decision Logic}\\
\vspace{0.2em}
Identify clinically appropriate candidate options
};

\node[dashedbox, fill=yellow!12, above=of clinical] (governance) {
\textbf{Layer 2: Governance Logic}\\
\vspace{0.3em}
Safety constraints\\
Stewardship constraints\\
Input completeness\\
Exclusion rules\\[0.3em]
\emph{Is it permissible to issue a recommendation?}
};

\node[box, fill=gray!15, above=of governance] (output) {
\textbf{System Output}\\
Recommendation \textbf{or} Abstention
};

\draw[arrow] (clinical) -- (governance);
\draw[arrow] (governance) -- (output);

\end{tikzpicture}

\caption{
Conceptual separation between clinical decision logic and governance within the proposed framework.
Clinical decision logic identifies candidate empiric options, while the governance layer determines whether
a recommendation may be issued given safety, stewardship, and information constraints.
The system output is explicitly limited to recommendation or abstention.
}
\label{fig:governance_architecture}
\end{figure}
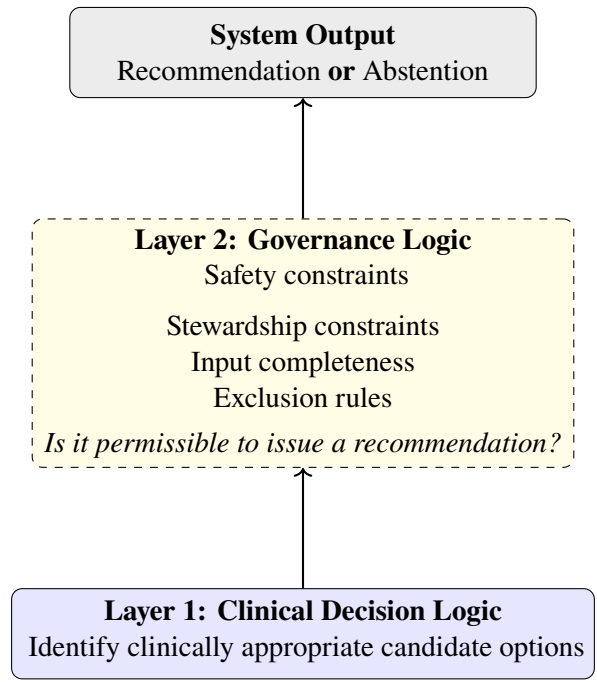

\FloatBarrier
\section{Governance and Safety Mechanisms}

Within the proposed framework, governance mechanisms are specified as explicit
control structures that determine whether a clinical recommendation may be issued
under a given set of inputs. Governance is treated as a first-class design component and
operates as a gatekeeping layer that constrains recommendation issuance.

Clinical decision rules define the conditions under which an empiric antibiotic class
may be considered appropriate within the defined scope. Governance mechanisms, in
contrast, evaluate whether recommendation issuance is permissible given information
completeness, uncertainty, exclusion criteria, and stewardship constraints.

Three core governance constructs are specified within the framework: a formal
abstention typology, deterministic stewardship constraints, and explicit exclusion rules.

\subsection{Abstention Typology}

Abstention is defined as an explicit system output indicating that no empiric antibiotic recommendation is issued. Within the framework, abstention is treated as a correct and intended outcome when predefined governance conditions are not satisfied, rather than as a residual or failure state.
A formal abstention typology is introduced to ensure that non-recommendation is interpretable, auditable, and reproducible. Abstention events are categorized into distinct, non-overlapping classes:
\begin{itemize}
\item \textit{Missing inputs}: Required clinical information necessary to evaluate recommendation eligibility is absent or incomplete.
\item \textit{Unknown or unquantified risk}: The clinical context involves risk factors that cannot be assessed within the defined scope or rule set.
\item \textit{Conflicting signals}:Available inputs activate mutually incompatible clinical or governance rules.
\item \textit{Explicit exclusions}: The case falls under a predefined exclusion criterion that prohibits recommendation issuance.
\item \textit{Conservative ambiguity}: Residual uncertainty persists despite permissive conditions, and conservative non-recommendation is preferred by design.
\end{itemize}

By making abstention categories explicit, the framework ensures that non-recommendation can be traced to a specific governance condition, supporting post hoc inspection and controlled system evolution.

\subsection{Deterministic Stewardship Constraints}

Antimicrobial stewardship is enforced through deterministic constraints that operate at
the governance level. Stewardship rules are specified as hard conditions that must be
satisfied for a recommendation to be issued and may veto otherwise permissible clinical
options.

These constraints include explicit preference for narrow-spectrum antibiotic classes
when clinically permissible, prohibition of unjustified escalation, and requirement for
explicit justification when broader-spectrum coverage is considered. Stewardship
constraints are applied uniformly and do not depend on heuristic weighting or
probabilistic scoring.

Stewardship enforcement is treated as a structural property of the decision-support
framework. Recommendations that violate predefined stewardship constraints are not
issued, even if they would otherwise satisfy clinical decision rules.

\subsection{Explicit Exclusion Rules}

Explicit exclusion rules define clinical contexts in which the system is specified to
abstain from recommendation by design. Exclusions are treated as first-class constructs
rather than as edge cases or implicit limitations.

Exclusion criteria may reflect contexts outside the defined scope, unresolved clinical
complexity, or conditions for which the framework is intentionally non-applicable.
When an exclusion rule is triggered, abstention is issued with an explicit exclusion
label, and no attempt is made to generate a recommendation.

This explicit handling of exclusions prevents silent generalization beyond validated
policies and supports transparent communication of system boundaries.

\section{Case-Based Evaluation Framework}

Evaluation within the proposed framework is designed to inspect system behavior rather
than to assess clinical effectiveness or predictive performance. The objective of
evaluation is to verify whether the system behaves in accordance with its explicitly
specified scope, governance rules, and design constraints.

Evaluation is conducted using a fixed set of synthetic clinical cases specifically
constructed to exercise distinct decision mechanisms. These cases are not intended to
represent real patient trajectories or epidemiological distributions. Instead, each case is
designed to probe a particular aspect of system behavior, such as abstention handling,
stewardship constraint enforcement, exclusion triggering, or recommendation eligibility
under permissive conditions.

Each case defines an expected system behavior that serves as the reference standard for
evaluation. Expected behavior is specified in terms of whether the system should issue
an empiric antibiotic class recommendation or abstain from recommendation, and, when
applicable, which antibiotic class is permissible within the defined policies. No
assessment of clinical outcomes or predictive accuracy is performed.

Within the evaluation framework, abstention aligned with predefined governance
conditions is treated as a behaviorally correct outcome. Recommendation issuance in
contexts that violate scope, governance, or stewardship constraints is considered
misalignment with expected behavior.

Case design follows a mechanism-driven approach. Rather than varying multiple factors
simultaneously, cases are constructed to isolate specific decision pathways, enabling
transparent attribution of system outputs to particular rules or constraints. This approach
supports reproducible inspection and facilitates targeted analysis of system behavior
under controlled conditions.

The case set used for evaluation is fixed and versioned as part of the framework
specification. All evaluation results are obtained by executing the same deterministic
evaluation procedure over the predefined cases, ensuring reproducibility and eliminating
dependence on sampling variability or stochastic effects.

\section{Evaluation Metrics}
Evaluation metrics within the proposed framework are descriptive and behavior-focused. They are designed to characterize how the system behaves under predefined
conditions, rather than to quantify performance, accuracy, or clinical effectiveness.

The primary metric reported is expected-behavior concordance, defined as the
proportion of cases in which the system output matches the predefined expected
behavior specified for each case. Concordance is assessed at the level of
recommendation versus abstention, and, when applicable, at the level of permissible
antibiotic class.

Recommendation coverage is reported as a complementary descriptive measure.
Coverage reflects the proportion of cases for which the system issues a
recommendation, stratified by case mechanism or governance condition. Coverage is
not interpreted as a measure of system completeness or desirability, but as a reflection
of the framework's conservative design and explicit abstention behavior.

Abstention reason distribution is reported to provide transparency into the governance
mechanisms that drive non-recommendation. Abstention events are categorized
according to the predefined abstention typology, enabling inspection of whether
governance conditions are activated as intended.

Stewardship-aligned behavior is assessed qualitatively by inspecting whether issued
recommendations adhere to predefined stewardship constraints. This includes
verification of narrow-spectrum preference when clinically permissible, absence of
unjustified escalation, and explicit justification when broader-spectrum coverage is
considered.
Determinism and reproducibility are assessed implicitly through repeated execution of
the evaluation procedure over the same fixed case set. Identical outputs across runs are
expected by design.

No statistical inference, hypothesis testing, or outcome-level evaluation is performed.
All reported metrics are intended solely to support transparent inspection of system
behavior within the explicitly defined scope.

\section{Reproducibility}

Reproducibility is a core design objective of the proposed framework. All aspects of
system behavior are fully determined by explicitly specified rules, a fixed set of
synthetic cases, and a deterministic evaluation procedure.

The evaluation artifacts consist of three primary components: (i) structured case
definitions encoded in a machine-readable format, (ii) an explicit evaluation framework
specifying expected behavior for each case, and (iii) a deterministic evaluation script
that applies the framework to the case set.

Case definitions are provided as structured data files that enumerate all input elements
required by the system. Each case includes a predefined expected behavior, specifying
whether a recommendation or abstention is anticipated and, when applicable, which
antibiotic class is permissible under the defined policies.

The evaluation script executes the same sequence of operations for every case, without
stochastic elements, random initialization, or adaptive logic. Repeated execution of the
script over the same case set yields identical outputs by design.

All reported observations of system behavior are obtained by executing the evaluation
script on the provided case set using the specified framework configuration. No manual
intervention, post hoc adjustment, or selective reporting is performed.

This explicit separation between system specification, case definition, and evaluation
execution enables independent reproduction of all reported observations. Any deviation
in system behavior can be traced to changes in the rule set, the case definitions, or the
evaluation procedure.

\section{Discussion}

This work presents a governance and evaluation framework for deterministic clinical
decision-support systems operating under explicitly constrained scope. The framework
addresses a specific methodological gap: the lack of explicit governance structures and
behavioral evaluation mechanisms in rule-based clinical decision support for high-stakes
prescribing contexts.

\subsection{Contributions and Design Principles}

A central contribution of the framework is the explicit separation between clinical
decision logic and governance mechanisms. This separation supports conservative
system behavior and enables traceable decision-support operation. Unlike approaches
that treat safety constraints as post-hoc filters or implicit heuristics, this framework
elevates governance to a first-class architectural component with explicit specification
and independent evaluability.

The formalization of abstention as an explicit and intended system outcome
distinguishes this framework from approaches that prioritize recommendation coverage.
By treating non-recommendation as a correct output when governance conditions are
not satisfied, the framework avoids the pressure to extrapolate beyond validated
policies—a critical consideration in high-stakes clinical contexts.

\subsection{Positioning Relative to Alternative Approaches}

The deterministic, rule-based design adopted here reflects specific trade-offs relative to
machine learning and probabilistic approaches:

\begin{itemize}
    \item \textbf{Explainability}: Every recommendation or abstention can be traced to explicit
    rule activation, whereas learned models may offer predictions without transparent
    decision pathways.
    \item \textbf{Scope control}: Deterministic systems operate strictly within defined boundaries,
    while statistical models may generalize beyond training distributions in unpredictable
    ways.
    \item \textbf{Validation requirements}: Behavioral alignment can be verified through
    mechanism-driven synthetic cases, whereas outcome-based validation requires
    clinical data and prospective evaluation.
    \item \textbf{Maintenance burden}: Rule-based systems require explicit updates as clinical
    policies evolve, whereas learned models may require retraining and revalidation.
\end{itemize}

These trade-offs suggest that deterministic frameworks are particularly appropriate
when:

\begin{itemize}
    \item Decision support must align with established clinical guidelines or institutional
    policies.
    \item Recommendation justification and auditability are regulatory or institutional
    requirements.
    \item The risk of inappropriate extrapolation outweighs the benefit of broader coverage.
    \item Clinical expertise is available for rule specification and maintenance.
\end{itemize}

Conversely, machine learning approaches may be more suitable when coverage
maximization is prioritized, when large-scale outcome data are available for training
and validation, or when decision policies are less well-defined.

\subsection{Methodological Implications}

The case-based evaluation strategy reflects a methodological choice to focus on
behavioral inspection rather than empirical performance assessment. Synthetic,
mechanism-driven cases allow controlled examination of specific decision pathways
without combining system behavior with clinical outcomes, population-level effects, or
real-world prevalence.

This evaluation approach is fundamentally different from outcome-based validation. It
does not assess whether the system improves clinical outcomes, reduces prescribing
errors, or enhances antimicrobial stewardship in practice. Instead, it verifies that the
system behaves as specified under controlled conditions—a necessary but not sufficient
condition for clinical deployment.

\subsection{Scope and Applicability}

The framework is intentionally narrow in scope. The reference context used for
illustration is limited to a specific empiric antibiotic prescribing scenario, and the
framework does not claim generalizability beyond this explicitly defined domain. This
narrow scope enables precise specification of governance rules and clinical logic with
domain-specific semantics. Clinical decision-support systems specifically targeting empirical antibiotic prescribing and antimicrobial stewardship have been developed and evaluated in various institutional contexts \cite{Akhloufi2022,Schaut2022}.

Broader applicability would require:

\begin{itemize}
    \item \textbf{Domain-specific adaptation}: Clinical rules, governance constraints, and
    stewardship requirements must be redefined for each new clinical context.
    \item \textbf{Scope validation}: The appropriateness of deterministic, rule-based decision
    support must be assessed for each target domain.
    \item \textbf{Governance recalibration}: Abstention typologies and exclusion criteria must
    reflect the uncertainty profile and safety requirements of each context.
\end{itemize}

The framework is positioned as a methodological template rather than a generalizable
system architecture.

\section{Limitations}

Several important limitations are explicitly acknowledged as part of the proposed
framework’s design philosophy. These limitations are not incidental, but rather reflect
deliberate choices made to prioritize safety, transparency, and controlled system
behavior.

First, the proposed framework does not provide clinical validation or outcome-based
evaluation. No real patient data are used, and no claims are made regarding clinical effectiveness, diagnostic accuracy, or impact on patient outcomes. The evaluation
framework is limited to behavioral inspection against predefined expected behavior
using synthetic cases.

Second, the framework is intentionally deterministic and rule-based. It does not
incorporate learning mechanisms, probabilistic inference, or adaptive behavior. As a
result, system behavior is fully dependent on the completeness, correctness, and scope
of the explicitly specified rules and governance constraints.

Third, the scope of applicability is narrowly defined. The framework is not designed to
generalize and no conclusions should be drawn regarding other clinical conditions,
populations, or care environments.

Fourth, the synthetic case set used for evaluation is mechanism-driven rather than
representative of real world clinical distributions. While this design supports controlled
inspection of system behavior, it does not capture epidemiological variability, workflow
complexity, or contextual factors present in clinical practice.

Finally, the framework does not address implementation, deployment, or integration
considerations. Issues related to user interaction, clinical workflow integration,
regulatory compliance, and real-world adoption are outside the scope of this work.

These limitations collectively underscore the methodological nature of the contribution.
The framework is intended as a specification and evaluation construct rather than as a
deployable clinical system or a substitute for clinical judgment.

\section{Conclusion}

This paper specifies a governance and evaluation framework for deterministic clinical
decision-support systems operating under explicitly constrained scope. By formalizing
governance as a first-class system component and treating abstention as a correct and
intended outcome, the framework provides a structured approach for developing
conservative, auditable decision support in high-stakes prescribing contexts.

The contribution is methodological: it demonstrates how explicit specification of scope,
governance mechanisms, and expected behavior can support disciplined development
and assessment of rule-based decision-support systems. The framework does not claim
clinical validation or readiness for deployment, but rather offers a reproducible template
for system specification, behavioral inspection, and governance enforcement.

In contexts where clinical decision support must align with established policies, where
recommendation justification is required, and where uncontrolled system behavior
poses safety risks, the explicit and conservative approach formalized in this framework
may provide a foundation for more transparent and traceable decision-support
development.

\bibliographystyle{elsarticle-num}

\typeout{get arXiv to do 4 passes: Label(s) may have changed. Rerun}
\end{document}